%% file: main.tex
\documentclass[a4paper,11pt]{article}

\usepackage{pos}
\usepackage{graphicx} 
\usepackage{url, hyperref}

\newcommand{\power}[2]{$#1 \times 10^{#2}$}

\title{Search for GeV-PeV neutrinos from nova T Coronae Borealis with IceCube}

\ShortTitle{T CrB neutrino search with IceCube}

\author{The IceCube Collaboration \\{\normalsize \normalfont(a complete list of authors can be found at the end of the proceedings)}\\}

\emailAdd{jthwaites@icecube.wisc.edu}
\emailAdd{justin.vandenbroucke@wisc.edu}

\abstract{
The widely anticipated outburst of recurrent nova T Coronae Borealis (T CrB), which is near the end of its 80-year cycle, provides an excellent opportunity to search for neutrinos from novae. Novae are an energetic class of transients, which have been studied for hundreds of years. Because many of them are located nearby, novae provide an excellent astrophysical laboratory to study shock-powered emission in our own backyard. Several recent novae have previously been detected in GeV gamma rays, and the 2021 outburst of RS Ophiuchi was detected up to TeV energies, with evidence for a hadronic origin of the observed emission. Previous searches for GeV-TeV neutrinos from novae, predicted to occur alongside their gamma-ray emission, have been performed using data from the IceCube Neutrino Observatory. However, no significant neutrino signals from novae have yet been observed. We present plans for follow-up of T CrB in real time with IceCube, using datasets spanning GeV to PeV neutrino energies.  Due to its closer distance and higher optical flux, which has been well measured in two historical eruptions, the expected neutrino signal from T CrB is several times stronger than that from RS Ophiuchi.  Furthermore, T CrB is located in the Northern sky at a declination where IceCube's sensitivity is an additional factor of a few better than at the location of RS Ophiuchi, which is beneficial to this search.

\vspace{4mm}

{\bfseries Corresponding authors:}
Jessie Thwaites$^{1}$,
Justin Vandenbroucke$^{1*}$\\
{$^{1}$ \itshape Department of Physics and Wisconsin IceCube Particle Astrophysics Center, University of Wisconsin—Madison}\\[4mm]
$^*$ Presenter
}

\input{ICRCdetails.tex}

\begin{document}

\maketitle

\section{Introduction}

Although novae are one of the longest known classes of astrophysical transients, they continue to provide new surprises thanks to modern instruments.  Novae occur when a white dwarf accretes matter from its stellar companion in a binary system, triggering a thermonuclear runaway explosion of the excess matter accumulated on the white dwarf surface (for a review, see~\cite{chomiuk_nova_review}).  Fermi-LAT made the surprise discovery that many novae emit GeV gamma rays simultaneously with their long-studied optical outbursts~\cite{Franckowiak:2017iwj}. Indeed, the LAT has detected the optically brightest (and generally closest) novae that have occurred since Fermi's launch.  This indicates that the gamma-ray flux is correlated with the optical flux, that the novae the LAT has not detected are simply below its detection threshold, and that likely all novae are gamma-ray sources.  In addition to the optical-gamma-ray flux correlation among novae, single novae exhibit correlated variability between the optical and gamma-ray light curves~\cite{aydi_v906car}. Because their optical emission dominates the bolometric luminosity, the observed correlation supports the theory that nova outbursts power hadronic shocks and that the shocks power not only the cosmic-ray acceleration resulting in gamma-ray emission, but also the optical outburst~\cite{Li:2017crr}.

In the hadronic shock scenario, gamma rays are produced by the decay of neutral pions.  Charged pions are produced along with neutral pions and decay producing neutrinos.  There is therefore growing evidence that nova outbursts are likely neutrino sources~\cite{bednarek:2022, magic_rso}. 

\begin{figure}[b!]
    \centering
    \includegraphics[width=0.7\linewidth]{ 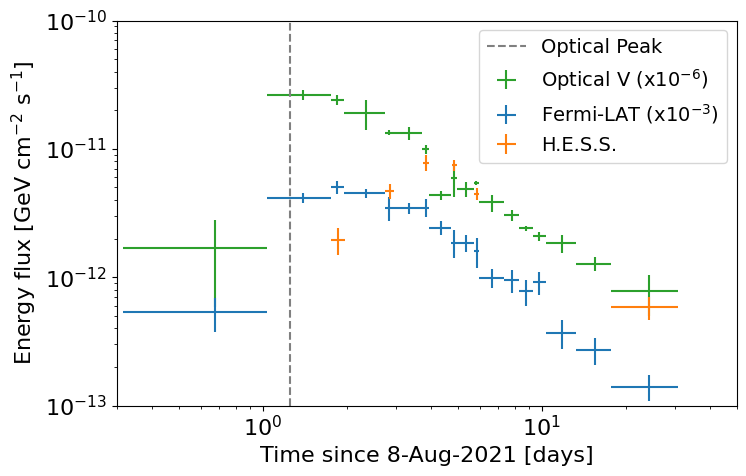}
    \caption{Light curve for the 2021 outburst of RS Oph in optical (V-band,~\cite{aavso}) and gamma rays (Fermi-LAT and H.E.S.S.,~\cite{hess_rso}). V-band data are averaged in a given bin, with bins chosen to match those in~\cite{hess_rso}. Fermi-LAT and optical data are scaled by the factors given in the legend, so that all three light curves can be shown compactly.}
    \label{fig:rso_lc}
\end{figure}

The 2021 outburst of nova RS Ophiuchi (RS Oph), a particularly nearby and bright recurrent nova, was detected not only in high-energy gamma rays by Fermi-LAT, but also in very-high-energy gamma rays by H.E.S.S.~\cite{hess_rso}, MAGIC~\cite{magic_rso}, and LST-1~\cite{lst1_rso}.  Some interpretations of these observations provide further support for the hadronic shock scenario.  Figure~\ref{fig:rso_lc} shows the optical and gamma-ray light curves of RS Oph.

The IceCube Neutrino Observatory consists of a billion tons of ice at the geographic South Pole instrumented with digital optical modules to record the optical Cherenkov light produced when neutrinos interact in the ice.  While the main IceCube array is optimized for TeV-PeV neutrinos, IceCube's dense infill array, IceCube DeepCore, provides excellent sensitivity in the 10 GeV to 1 TeV energy range.  This is the range in which nova gamma rays have been detected and neutrinos are expected.

An event selection, named GeV Reconstructed Events with Containment for Oscillation (GRECO), was initially developed for neutrino oscillation physics with DeepCore.  It was subsequently adapted to GRECO Astronomy for sub-TeV astrophysics measurements with DeepCore (see Appendix B of~\cite{novae_icecube} for a complete description).  Using GRECO Astronomy, we have previously searched for neutrinos from both optically selected and gamma-ray-selected novae~\cite{novae_icecube}.  We subsequently performed a dedicated search for neutrinos from the 2021 RS Oph outburst~\cite{rso_proceeding}.  All of these analyses resulted in null detections and upper limits.  A recent paper~\cite{PRD_modeling:2025} modeled MeV and $>$$1$~GeV neutrinos from novae including T CrB and compared the predicted spectrum to the sensitivity of the main IceCube array; unfortunately it neglected to include IceCube DeepCore, which provides enhanced sensitivity in the GeV-TeV energy range relevant to novae.

Historic eruptions from a nova even closer than RS Oph, named T Coronae Borealis (T CrB), were detected optically in 1886 and again in 1946.  Like many novae, this is a ``recurrent'' nova, i.e., one that has been detected multiple times.  Although they do not follow precise periodicity, the recurrent novae do recur with approximately the same time between outbursts.  This is believed to be the timescale over which the white dwarf accretes outer shell material before it accumulates sufficient matter to trigger thermonuclear runaway, expelling the shell and starting the process again.  Novae that have not been observed to be recurrent may simply have long recurrence times.  Based on the historic eruption dates, T CrB is expected to erupt in approximately 2026.  The historic observations indicate a pre-eruption dimming, or ``dip'', in the light curve.  Some authors have used this feature to attempt a more precise prediction of the eruption time, resulting in an estimate of an outburst in 2024~\cite{Schaefer_prediction}.  Although precisely predicting the T CrB eruption is challenging, it is likely to occur at any time now.

\begin{figure}
    \centering
    \includegraphics[width=0.7\linewidth]{ 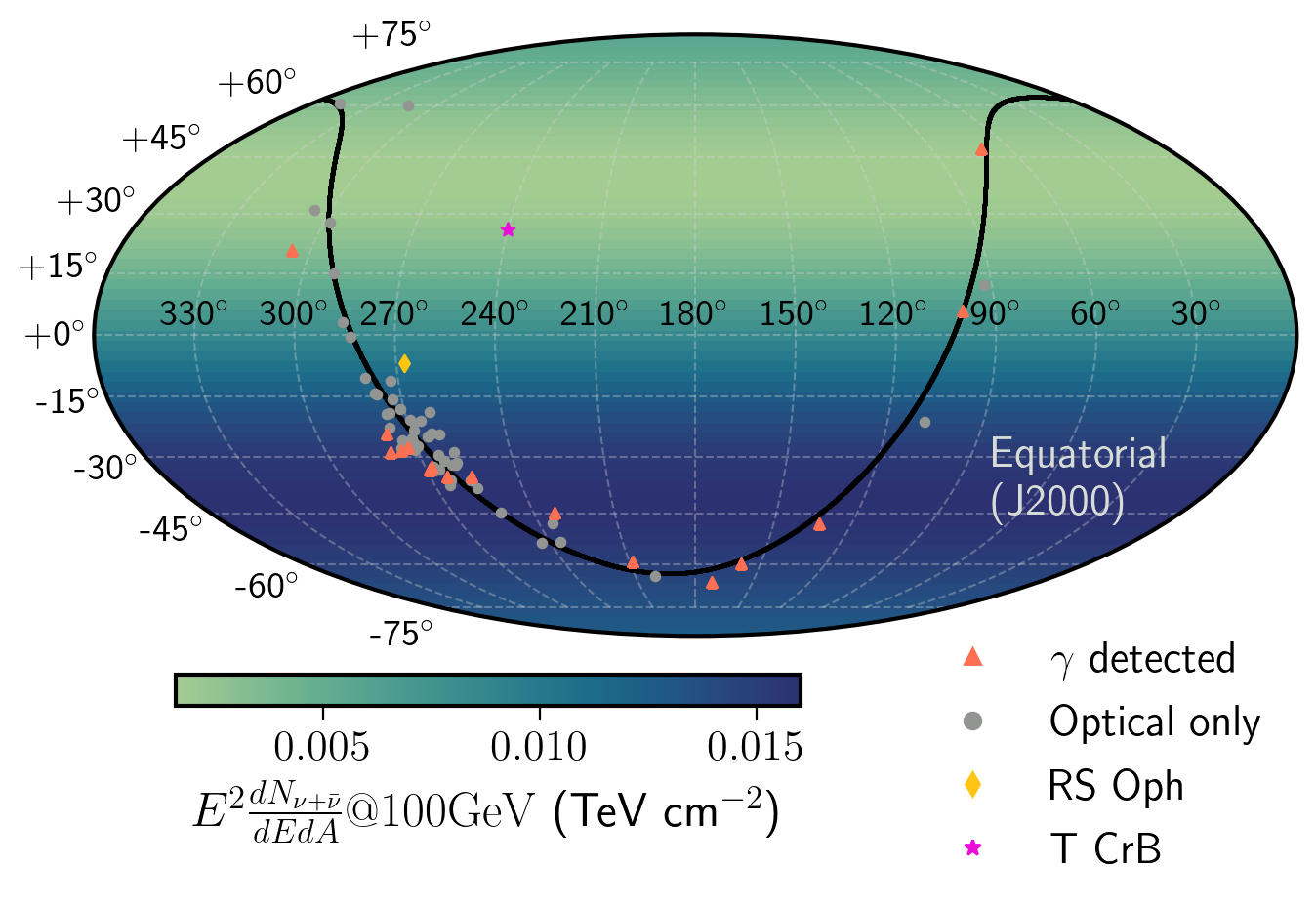}
    \caption{Skymap showing the location of novae studied with IceCube. Novae previously analyzed with IceCube are shown as gray circles for those only optically detected, while those also detected in gamma rays are shown in orange triangles.  The background color indicates the GRECO Astronomy sensitivity to a source with spectral index 2.0 for a time window of one day~\cite{novae_icecube}. The locations of nova RS Oph (yellow diamond) and T CrB (purple star) are also indicated.}
    \label{fig:skymap}
\end{figure}

Based on previous optical and gamma-ray observations as well as our own previous studies of novae including RS Oph, we present predictions of the expected neutrino signal.  We furthermore use the expected duration of the upcoming outburst to optimize our planned search for neutrinos.  In addition to being closer, with a brighter expected neutrino flux, T CrB is at a more favorable declination for the IceCube DeepCore (GRECO Astronomy) sensitivity, shown in Figure~\ref{fig:skymap}.  Having optimized the analysis, we plan to perform it rapidly when the outburst occurs, in order to provide notification of a detection or non-detection in near real time to other observers.

\section{Sensitivity and Neutrino Detection Prospects}

For the expected upcoming outburst of T CrB, we plan to use two analyses previously applied to search for neutrino emission from novae in real time. The analyses cover two complementary energy ranges: (1) GeV-TeV energies using GRECO Astronomy~\cite{novae_icecube}, and (2) TeV-PeV energies using the Fast Response Analysis (FRA), as was done with RS Oph~\cite{rso_atel}.

\begin{figure}
    \centering
    \includegraphics[width=0.75\linewidth]{ 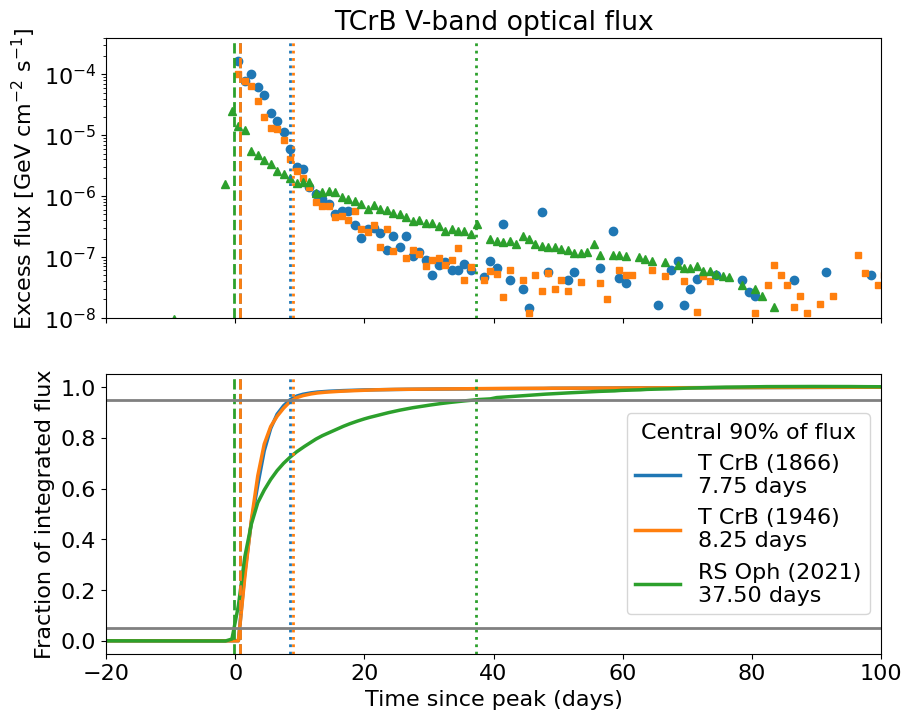}
    \caption{Calculation of the time to accumulate the central 90\% of the flux during the outburst for historical T CrB and recent RS Oph eruptions. The top panel shows the lightcurve with the baseline flux subtracted in order to calculate the excess flux during the outburst. The lightcurve is calculated by binning the data in 1 day bins and averaging in each bin. The lightcurve is then integrated, with the fraction of integrated flux accumulated shown in the bottom panel. 
    The beginning (dashed) and end (dotted) of the $T_{90}$ interval are shown as vertical lines.}
    \label{fig:lightcurves}
\end{figure}

To prepare for a search for GeV-TeV neutrinos from nova T CrB with IceCube, we optimize the analysis based on the optical light curve of the two previous outbursts of this nova. To quantify the outburst duration, we calculate $T_{90}$, defined as the time duration when $5\%-95\%$ of the total flux is accumulated (as is commonly defined for gamma-ray bursts), using optical V-band data from previous outbursts of both RS Oph and T CrB. This calculation, as well as the complete light curve of the nova, is shown in Figure~\ref{fig:lightcurves} and summarized in Table~\ref{tab:rso_tcrb_params}.

\begin{table}[]
    \centering
    \begin{tabular}{|c|c|c|c|c|}
        \hline
        Outburst & Peak Magnitude & Peak Flux & Integrated Flux &$T_{90}$ Duration \\ 
         & [mags] & [GeV cm$^{-2}$ s$^{-1}$] & [GeV cm$^{-2}$] & [Days]\\ 
        \hline\hline 
        RS Oph 2021 & 4.5 & \power{3.23}{-5} & 8.87 & 37.50 \\ 
        \hline 
        T CrB 1866 & 2.0 & \power{3.27}{-4} & 52.32 & 7.75 \\ 
         & & = 10.12 $\times$ RS Oph & =5.90 $\times$ RS Oph & \\ 
        \hline
        T CrB 1946 & 3.0 & \power{1.30}{-4} & 35.09 & 8.25 \\ 
         & & = 4.03 $\times$ RS Oph & =3.95 $\times$ RS Oph & \\ 
        \hline
    \end{tabular}
    \caption{V-band optical light curve parameters for RS Oph~\cite{aavso} and T CrB~\cite{Schaefer_lc}. The optical flux is calculated by converting to flux density in Janskys and then multiplying by the FWHM of the Johnson V-band. The flux is averaged in six-hour bins and numerically integrated to determine the integrated flux.}
    \label{tab:rso_tcrb_params}
\end{table}

\begin{figure}
    \centering
    \includegraphics[width=0.49\linewidth]{ 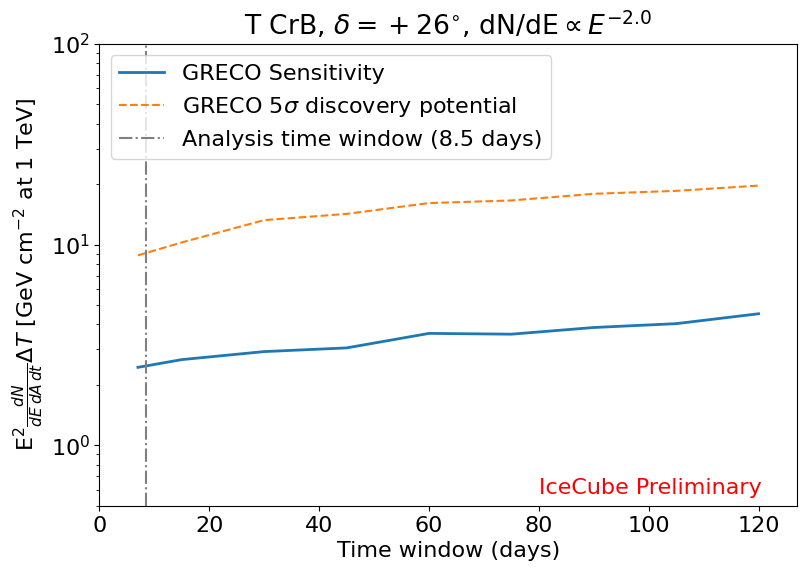}
    \includegraphics[width=0.48\linewidth]{ 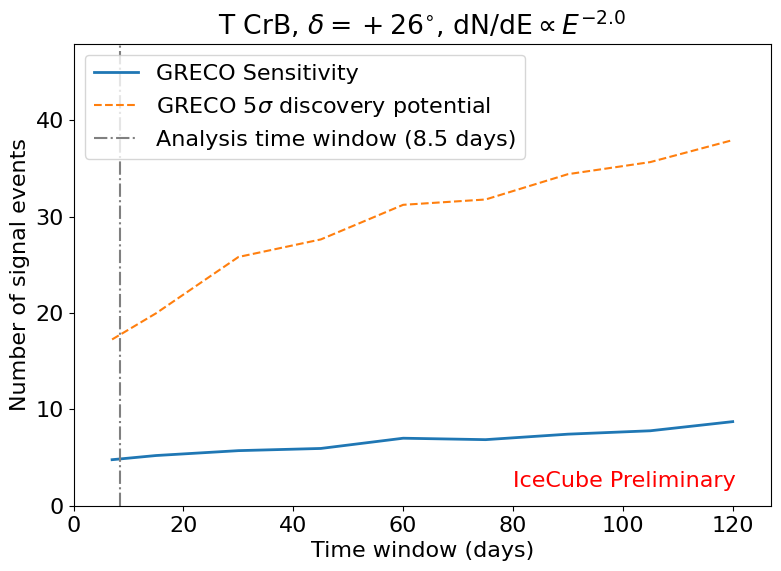}
    \caption{Time-integrated per-flavor flux sensitivity and $5\sigma$ discovery potential versus time window for a source at the T CrB declination ($+26^\circ$) using GRECO Astronomy, for a source with particle flux $dN/dE \propto E^{-2}$. The left panel shows the time-integrated sensitivity, while the right panel shows the sensitivity in number of signal neutrino events.}
    \label{fig:sensitivity}
\end{figure}

For the IceCube search we plan a time window of $[-0.5,\, +8.0]$ days relative to the optical peak, in order to cover the $T_{90}$ time and possible early or late emission. The sensitivity and $5\sigma$ discovery potential at this declination as a function of the analysis time window are shown in Figure~\ref{fig:sensitivity}. The short duration of this nova compared to RS Oph is advantageous for neutrino follow-up, as it reduces the expected number of background events in the direction of the nova.

\begin{figure}
    \centering
    \includegraphics[width=0.6\linewidth]{ 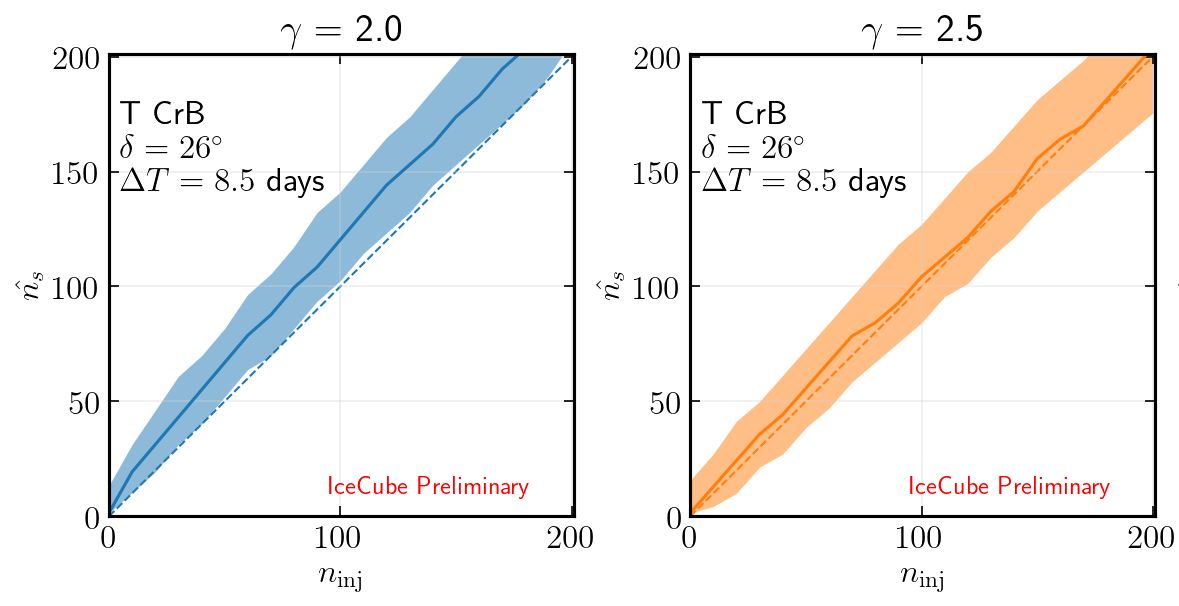}
    \includegraphics[width=0.37\linewidth]{ 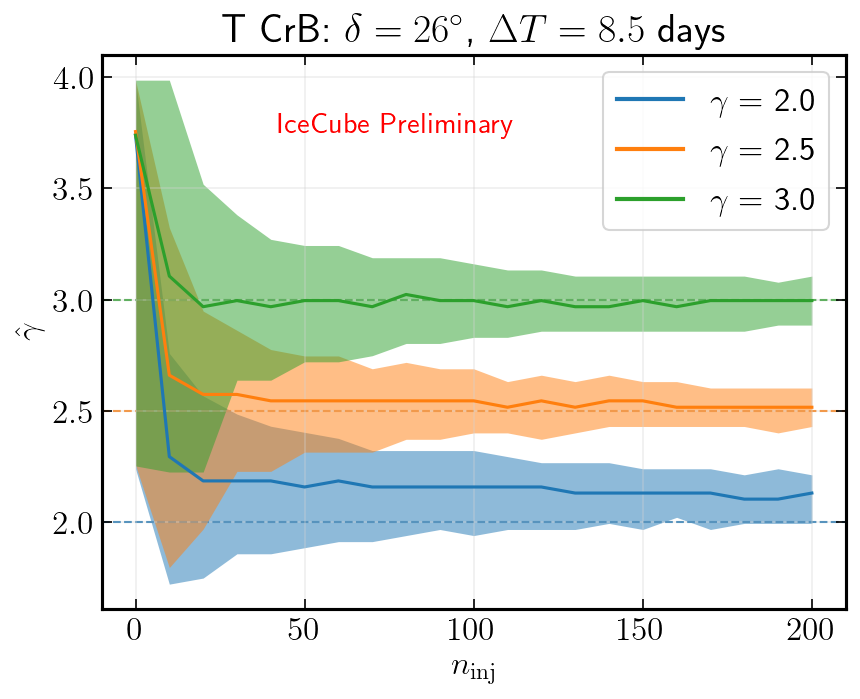}
    \caption{Ability of the GRECO Astronomy nova analysis to recover injected signal, for an analysis time window of 8.5 days. The left (middle) panel shows the fitted number of signal events for a true spectral index of 2.0 (2.5). The right panel shows the fitted spectral index for three injected spectral indices: 2.0, 2.5, and 3.0. Each band indicates the $\pm1\sigma$ variation.}
    \label{fig:bias}
\end{figure}

We also assess the ability of this sub-TeV analysis to recover the simulated signal for the 8.5-day time window, as shown in Figure~\ref{fig:bias}. For a hard spectral index ($dN/dE\propto E^{-2.0}$), the analysis tends to fit a somewhat softer spectral index and slightly overestimate the number of signal events, while for softer indices ($dN/dE\propto E^{-2.5}$ or $E^{-3.0}$) the analysis recovers both the injected number of signal events and the spectral index with negligible bias.

\begin{figure}
    \centering
    \includegraphics[width=0.49\linewidth]{ 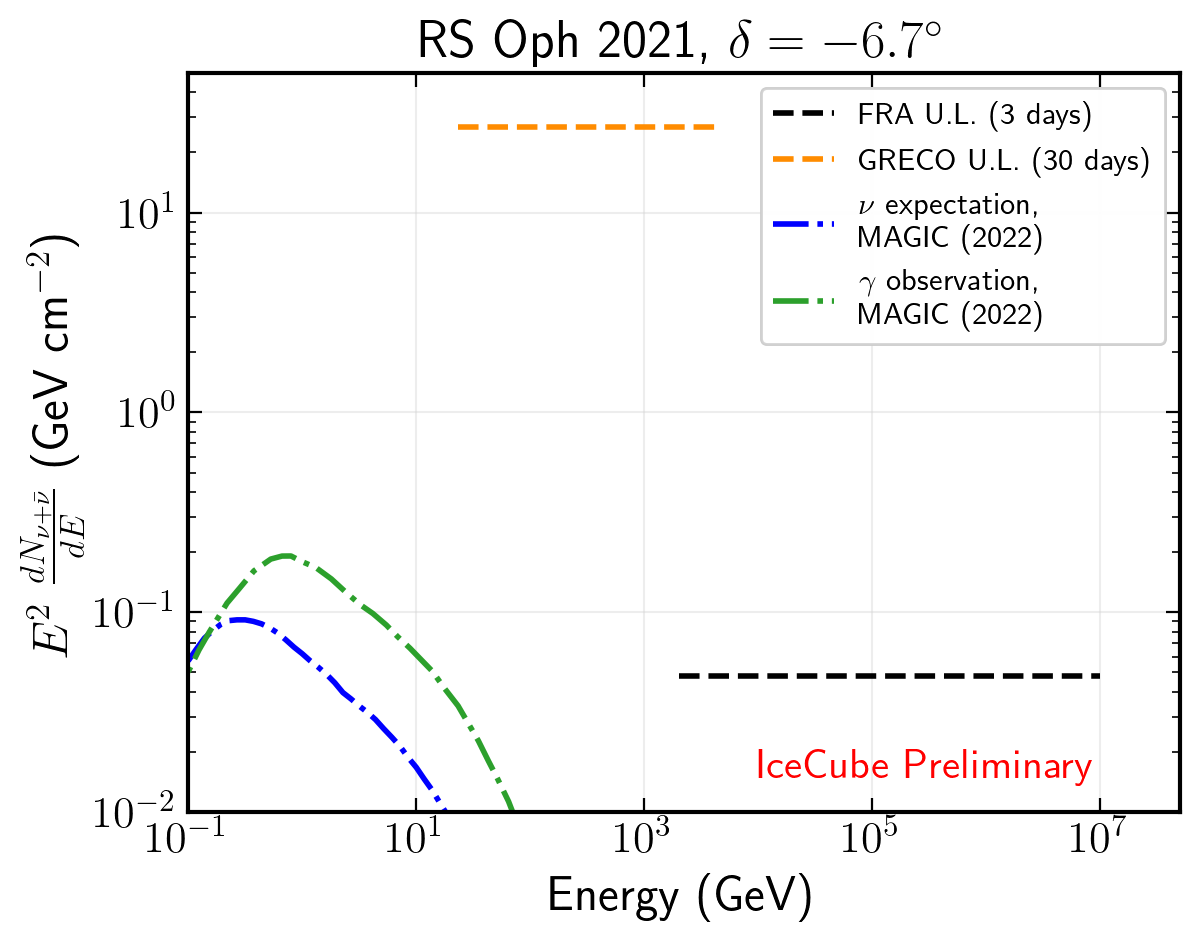}
    \includegraphics[width=0.49\linewidth]{ 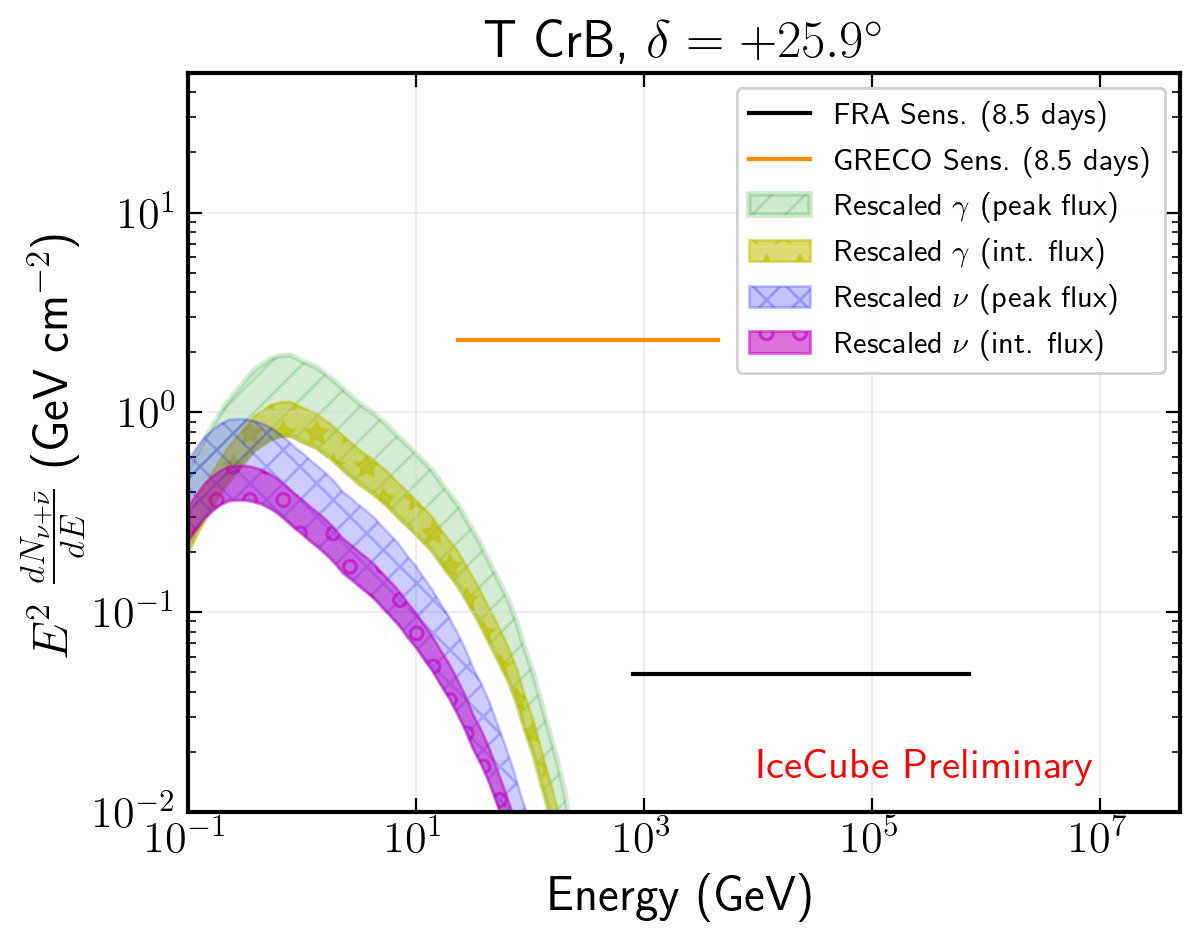}
    \caption{
    \textit{Left:} Per-flavor upper limits for RS Oph's 2021 outburst with the Fast Response Analysis (FRA, using a high-energy IceCube event selection) and GRECO Astronomy dataset. The green dot-dashed curve shows a hadronic model fit to observations by MAGIC~\cite{magic_rso}, over the four days of observation averaged in the model fit. The $\nu_\mu+\nu_e$ neutrino expectation from the hadronic model calculated in~\cite{magic_rso}, after accounting for oscillations (assuming 1:1:1 flux at Earth and shown as per-flavor fluxes), is plotted as a blue dot-dashed curve.
    \textit{Right:} Per-flavor sensitivity for T CrB with FRA and GRECO Astronomy in an 8.5-day time window. Two possible neutrino and gamma-ray spectrum expectations are calculated for T CrB, by rescaling the gamma-ray observations and neutrino expectation from RS Oph by a ratio of the optical fluxes from the two novae using values in Table~\ref{tab:rso_tcrb_params} (see text for details). The energy range of each sensitivity line corresponds to the central $90\%$ range.}
    \label{fig:nu_exp}
\end{figure}

In order to estimate the expected neutrino flux from T CrB, we compare the V-band optical data from its two historic outbursts to the 2021 RS Oph outburst (see Table~\ref{tab:rso_tcrb_params}). We begin with the hadronic model fit to the GeV-TeV flux, along with the neutrino flux predicted by this hadronic model, provided by~\cite{magic_rso}. These spectra, after accounting for neutrino oscillations, and the IceCube upper limits on neutrino emission from RS Oph, are shown in Figure~\ref{fig:nu_exp} (left panel). For a scenario in which the neutrino and gamma-ray fluxes scale with the peak optical flux, we rescale the neutrino and gamma-ray curves by the ratio of the optical peak flux from the two historical outbursts to that from RS Oph 2021. For an alternative scenario in which the fluxes scale with the time-integrated flux over the outburst, we rescale by the ratio of the total integrated optical flux in the two historical outbursts to that of RS Oph. Both of these estimates are shown in Figure~\ref{fig:nu_exp} (right panel). In addition, the location of T CrB in the Northern sky (see Figure~\ref{fig:skymap}) and the shorter outburst duration improve IceCube's sensitivity to this nova, improving the prospects for detecting it in GeV-TeV neutrinos.

\section{Conclusion and Outlook}

While DeepCore, including the GRECO Astronomy event selection, has been used previously for several archival astrophysical neutrino searches, it has not yet been used for real-time neutrino astrophysics.  Motivated especially by the upcoming T CrB outburst, we have performed the technical improvements required to make the GRECO Astronomy data available in low latency.  This is necessary preparation for T CrB and also lays the groundwork for other real-time sub-TeV astrophysical neutrino searches with DeepCore and the IceCube Upgrade.

Based on the historical eruptions of nova T CrB in 1866 and 1946, it is expected to erupt again in the near future.  This will likely be the closest nova outburst in a generation, resulting in the brightest flux in various wavebands including optical, high-energy gamma rays, and very-high-energy gamma rays.  Based on the ratio of optical to gamma-ray flux measured for RS Oph, we have estimated the expected gamma-ray and corresponding neutrino energy spectrum for T CrB.  Because the two historical T CrB eruptions exhibited remarkably similar optical light curves, it is likely that the upcoming outburst will have a similar light curve.  Assuming the correlation measured between optical and gamma-ray light curves in other novae also applies to T CrB, the historical optical data provide a good indication of the expected temporal evolution of the gamma-ray and neutrino signals in its upcoming eruption.

In addition to these signal predictions, we have calculated the sensitivity of IceCube (in particular DeepCore) to the upcoming outburst and optimized and validated the planned analysis using simulated signals.  T CrB is at a more favorable declination than RS Oph for IceCube, resulting in a sensitivity within an order of magnitude of our  signal prediction.  If the signal is brighter than estimated according to our simple scaling calculations, detection of neutrinos from T CrB is within IceCube's reach. These findings are consistent with theoretical calculations of the expected neutrino flux from novae~\cite{Partenheimer:2024qxw}.

The IceCube Upgrade is a project to further improve IceCube's sub-TeV neutrino sensitivity~\cite{Aya_upgrade}.  It consists of seven new strings, whose instrumentation has been produced and which will be installed in 2025-2026, potentially in time for the T CrB eruption.  The strings will provide an even denser infill within the DeepCore infill array.  The IceCube Upgrade will boost the IceCube effective area and energy and reconstruction capabilities in the 1-100 GeV energy range~\cite{Kobayashi:icrc2025}, a band well suited for novae including T CrB. The Upgrade will also provide precision calibration of the optical properties of the South Pole ice, resulting in improved IceCube performance across all energies.  Compared to the demonstrated performance of IceCube, including DeepCore, the IceCube Upgrade will soon extend our sensitivity to T CrB and other neutrino source candidates.

\bibliographystyle{ICRC}
\bibliography{references}

\clearpage
\input{authorlist_IceCube_with_AAVSO_ack}
\end{document}

%% file: ICRCdetails.tex
\ConferenceLogo{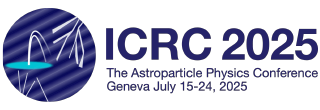}

\FullConference{39th International Cosmic Ray Conference (ICRC2025)\\
 15–24 July 2025\\
Geneva, Switzerland\\}

%% file: authorlist_IceCube_with_AAVSO_ack.tex
\section*{Full Author List: IceCube Collaboration}

\scriptsize
\noindent
R. Abbasi$^{16}$,
M. Ackermann$^{63}$,
J. Adams$^{17}$,
S. K. Agarwalla$^{39,\: {\rm a}}$,
J. A. Aguilar$^{10}$,
M. Ahlers$^{21}$,
J.M. Alameddine$^{22}$,
S. Ali$^{35}$,
N. M. Amin$^{43}$,
K. Andeen$^{41}$
% ...
\\
\\
% ...
$^{10}$ Universit{\'e} Libre de Bruxelles, Science Faculty CP230, B-1050 Brussels, Belgium \\
% ...
$^{16}$ Department of Physics, Loyola University Chicago, Chicago, IL 60660, USA \\
$^{17}$ Dept. of Physics and Astronomy, University of Canterbury, Private Bag 4800, Christchurch, New Zealand \\
% ...
$^{21}$ Niels Bohr Institute, University of Copenhagen, DK-2100 Copenhagen, Denmark \\
$^{22}$ Dept. of Physics, TU Dortmund University, D-44221 Dortmund, Germany \\
% ...
$^{35}$ Dept. of Physics and Astronomy, University of Kansas, Lawrence, KS 66045, USA \\
% ...
$^{39}$ Dept. of Physics and Wisconsin IceCube Particle Astrophysics Center, University of Wisconsin{\textemdash}Madison, Madison, WI 53706, USA \\
% ...
$^{41}$ Department of Physics, Marquette University, Milwaukee, WI 53201, USA \\
% ...
$^{43}$ Bartol Research Institute and Dept. of Physics and Astronomy, University of Delaware, Newark, DE 19716, USA \\
% ...
$^{63}$ Deutsches Elektronen-Synchrotron DESY, Platanenallee 6, D-15738 Zeuthen, Germany \\
$^{\rm a}$ also at Institute of Physics, Sachivalaya Marg, Sainik School Post, Bhubaneswar 751005, India \\
% ...

\subsection*{Acknowledgments}

\noindent
We acknowledge with thanks the variable star observations from the AAVSO International Database contributed by observers worldwide and used in this research.
The authors gratefully acknowledge the support from the following agencies and institutions:
USA {\textendash} U.S. National Science Foundation-Office of Polar Programs,
U.S. National Science Foundation-Physics Division,
U.S. National Science Foundation-EPSCoR,
U.S. National Science Foundation-Office of Advanced Cyberinfrastructure,
Wisconsin Alumni Research Foundation,
Center for High Throughput Computing (CHTC) at the University of Wisconsin{\textendash}Madison,
Open Science Grid (OSG),
Partnership to Advance Throughput Computing (PATh),
Advanced Cyberinfrastructure Coordination Ecosystem: Services {\&} Support (ACCESS),
Frontera and Ranch computing project at the Texas Advanced Computing Center,
U.S. Department of Energy-National Energy Research Scientific Computing Center,
Particle astrophysics research computing center at the University of Maryland,
Institute for Cyber-Enabled Research at Michigan State University,
Astroparticle physics computational facility at Marquette University,
NVIDIA Corporation,
and Google Cloud Platform;
Belgium {\textendash} Funds for Scientific Research (FRS-FNRS and FWO),
FWO Odysseus and Big Science programmes,
and Belgian Federal Science Policy Office (Belspo);
Germany {\textendash} Bundesministerium f{\"u}r Forschung, Technologie und Raumfahrt (BMFTR),
Deutsche Forschungsgemeinschaft (DFG),
Helmholtz Alliance for Astroparticle Physics (HAP),
Initiative and Networking Fund of the Helmholtz Association,
Deutsches Elektronen Synchrotron (DESY),
and High Performance Computing cluster of the RWTH Aachen;
Sweden {\textendash} Swedish Research Council,
Swedish Polar Research Secretariat,
Swedish National Infrastructure for Computing (SNIC),
and Knut and Alice Wallenberg Foundation;
European Union {\textendash} EGI Advanced Computing for research;
Australia {\textendash} Australian Research Council;
Canada {\textendash} Natural Sciences and Engineering Research Council of Canada,
Calcul Qu{\'e}bec, Compute Ontario, Canada Foundation for Innovation, WestGrid, and Digital Research Alliance of Canada;
Denmark {\textendash} Villum Fonden, Carlsberg Foundation, and European Commission;
New Zealand {\textendash} Marsden Fund;
Japan {\textendash} Japan Society for Promotion of Science (JSPS)
and Institute for Global Prominent Research (IGPR) of Chiba University;
Korea {\textendash} National Research Foundation of Korea (NRF);
Switzerland {\textendash} Swiss National Science Foundation (SNSF).